\documentclass[conference]{IEEEtran}
\IEEEoverridecommandlockouts
\usepackage{cite}
\usepackage{amsmath,amssymb,amsfonts}
\usepackage{algorithmic}
\usepackage{graphicx}
\graphicspath{{data/}{images/}}
\usepackage{textcomp}
\usepackage{listings}  
\usepackage{todonotes}
\usepackage{tabularx}
\usepackage{comment}
\usepackage{dirtytalk}
\usepackage{xcolor}
\usepackage{caption}
\usepackage{subcaption}
\usepackage{soul}
\usepackage[]{quoting}
\usepackage{multirow}

\usepackage[utf8]{inputenc}

\usepackage{booktabs}
\usepackage{prettyref}

\makeatletter
\def\ScaleIfNeeded{%
\ifdim\Gin@nat@width>\linewidth
\linewidth
\else
\Gin@nat@width
\fi
}
\makeatother

\begin{document}

\IEEEoverridecommandlockouts
\IEEEpubid{\makebox[\columnwidth]{\textbf{ 978-1-7281-6680-3/20/\$31.00~\copyright2020 IEEE }\hfill} \hspace{\columnsep}\makebox[\columnwidth]{  }}

\title{ADW: Blockchain-enabled Small-scale Farm Digitization}  
\makeatletter
\newcommand{\linebreakand}{%
  \end{@IEEEauthorhalign}
  \hfill\mbox{}\par
  \mbox{}\hfill\begin{@IEEEauthorhalign}
}
\makeatother

\author{\IEEEauthorblockN{Nelson Bore, Andrew Kinai, Peninah Waweru, Isaac Wambugu \\Juliet Mutahi, Everlyne Kemunto, Reginald Bryant, Komminist Weldemariam} 
\textit{IBM Research $|$ Africa} \\
  Nairobi, Kenya 

}

\maketitle
\IEEEpubidadjcol
\begin{abstract}
Farm records hold the static, temporal, and longitudinal details of the farms. For small-scale farming, the ability to accurately capture these records plays a critical role in formalizing and digitizing the agriculture industry. Reliable exchange of these record through a trusted platform could unlock critical and valuable insights to different stakeholders across the value chain in agriculture eco-system. Lately, there has been increasing attention on digitization of small scale farming with the objective of providing farm-level transparency, accountability, visibility, access to farm loans, etc. using these farm records. However, most solutions proposed so far have the shortcoming of providing detailed, reliable and trusted small-scale farm digitization information in real time. To address these challenges, we present a system, called Agribusiness Digital Wallet (ADW), which leverages blockchain to formalize the interactions and enable seamless data flow in small-scale farming ecosystem. Utilizing instrumentation of farm tractors, we demonstrate the ability to utilize farm activities to create trusted electronic field records (EFR) with automated valuable insights. Using ADW, we processed several thousands of small-scale farm-level activity events for which we also performed automated farm boundary detection of a number of farms in different geographies.

\end{abstract}


\section{Introduction} \label{sec:introduction}


Agricultural information exchange (AIE) is the electronic transfer of farm related activities and administrative information across the complex value chains with the objective of delivering the right data and information to the appropriate person at the right time. AIE is critical for the full realization of Electronic Field Record (EFR) benefits and has the potential for improving food \emph{security} while reducing post-harvest losses, facilitating trusted communication and data sharing among critical farm ecosystem participants to enable farm-level decision makings, supply and demand visibility and efficient, trusted, data-driven agriculture.

The use of EFRs could also improve the financial sector engagement by enabling banks to have quick visibility to farm records and increasing their interactions with financial service seeking behavior in real time. However, despite its benefits, the adoption of EFRs has been affected by various challenges \cite{FlemingKOW16-2,FlemingWWOS16}.
Moreover, the informal interactions of the small-scale farming ecosystem (for {\say{doing-agri-business}}) in developing countries contributes to three challenges: fragmented processes, provenance and trust, and information visibility.

We carried out an in-depth investigation of the challenges of small-scale farming by conducting design thinking sessions with stakeholders in the ecosystem. \textit{Fragmented processes} conveyed high transaction cost, complex, and ad-hoc (rather unstructured and informal) communication among the participants of the small-scale farming ecosystem  makes it very difficult to efficiently and transparently undertake agribusiness transactions across the value chain.  The \textit{provenance and trust} challenge in the ecosystem is attributed to data, lack of end-to-end data value chain across the network, equipment (e.g., tractors) provenance is not accurately known (including parts and maintenance history) and hence warranty of tractors cannot be justified as well as tractors are most likely maintained when they are out-of-service.  The \textit{information visibility} hinders informative decisions for the growth and productivity of small-scale farming in emerging market countries where achieving their \emph{food security} is tightly coupled with the success of small-scale farming. 

\IEEEpubidadjcol
\IEEEpubidadjcol

There are three opportunities for modernization and digitization in this domain that can greatly enhance the faster flow of goods, information, and money. First, the formalizing of farms themselves through digital twin representations can bring much needed source information into the decision making process. Second, traceability exists in agriculture ecosystems today in a very limited way, and bringing end-to-end visibility across all aspects of agriculture ecosystems can play a critical role in ensuring food safety and provenance. Third, monetary transactions like trading, credit, and insurance can benefit greatly with access to information coming from the first two making these important commercial activities streamlined, data based, and paperless.

This paper focuses on the first opportunity, i.e., on formalizing small-scale farming. In particular, we present a blockchain-based platform, which we call the Agribusiness Digital Wallet (ADW). ADW is designed to establish a digital trust among the agriculture stakeholders (farmers, co-ops/agents, tractor owners/dealers, financial institutions, input suppliers, etc.) and enable decision supports while unlocking values for the stakeholders of the value chains. 

The use of blockchain for agriculture and agribusiness has been discussed in a number of recent papers ---e.g.,\cite{8718621,8500647,8373021,8605963,8615007}. They predominately focused on aspects of tracking across the value/supply chains for achieving provenance and traceability of agricultural products. However, none of these attempt to tackle the problems of formalizing interactions among the stakeholders and data value chain of small-scale farming via digitizing.
Differently, our work primarily focused on
digitization of small-scale farming.


More specifically, the core contribution of ADW is providing: (i) a tamper-proof definition of demand-side and supply-side workflows from farmer tractor request to fulfilment, payment for services, and distribution of proceeds; (ii) authorized access to services and documents within the workflow; (iii) logging of all workflow-related approvals, and (iv) secure storage of all data, events and documents, including documented workflow (e.g., tractor booking, invoicing, etc.). 

We process and maintain internet of things(IoT) data (tractor generated) and transactions on the blockchain while enabling interoperability and the sharing of verified data and including documents and eventually facilitating sharing services (e.g., an advisor that can guide credit underwriting and tractor investment decisions in a given market) as the business network grows.  
We keep track of digital confirmation of receipt of events at different steps, 
documentation availability, tractor status, etc. We provide features to retrieve any of the above information as part of information visibility, including reporting on tractor and operator activity, requests for tractor services, the maintenance needs of a tractor, etc.  Finally, using the verified IoT data and machine learning models, we provide analytics services such as automated field boundary identification and acreage estimation, by coupling with other data sources such as weather and remote sensing data. All these can be accessed via the verified EFR for each farm.








\section{Background and Motivation}

In this section, we present the field studies and workshops we have conducted to understand and characterize pain points in small-scale farming. We then establish the necessary requirements for this work by synthesizing the pain points. 

\subsection{Challenges of small-scale farming ecosystem}

To gain a deeper understanding of the decision-making process among the stakeholders in the small-scale farming ecosystem, we conducted a 3-day long workshop (in Abuja, Nigeria). The participants of the workshop include representatives from farmers and farm cooperative networks, mechanization logistics companies, and representatives from financial institutions (3 banks who specialized in agribusiness financing participated). The workshop was centered around understanding the current pain points of the informal agriculture value chains of the small-scale farming ecosystem. 
In particular, first, we focused on understanding how new data sets can be captured for small-scale farming to mitigate any challenge that may hinder timely data-driven decision making. 
Second, we assessed how new and existing data sets can be enhanced into capabilities for formalizing and digitization the ecosystem to build a trusted electronic field record (EFR).  

During the workshop, each of the representative detailed  pain points they have been experiencing with over the years.  We noted the quality of information flow, farm data, and any farm-related activities (and farm services) depend on {the trust and credibility of farm extension agents/workers}, typically trained agronomists (they do have different names in different geographies such as field agents/officers, farm agents, etc.). This makes them the epicenter of data and any other farm-related activities. They assist in collecting (often manually and informally) details on farm size, farm profile, 
farm geo-location, expected yield, and type of service requested by farmers. Farm input providers who provide improved seeds, matching fertilizer, farm machinery services, etc. rely on these farm extension agents. The agents also serve as field supervisors that assess the quality of all the mechanization services done by tractors and thereby approve service payments to be made to tractor owners for the completed services. However, the trust and credibility of these agents and hence their services significantly degraded over time as suggested by all participants of the workshop. As a result of this, any reported farm data and farm activities about small-scale farming ecosystems cannot be used \say{as-is} for any upstream and downstream decision makings. Arguably, for our team, this can be termed as a \textbf{single point of failure}. This leads to a number of acute challenges for the full realization of EFR for small-scale farming ecosystem particularly in developing countries, including:
\begin{itemize}
\item Small-scale farmers continue to struggle receiving quality advisory services (including weather information for a timely decision on when, how and what to plant, etc.) to optimize crop yield.  
\item Aggregators continue to struggle to secure sufficient volumes of crops to meet their clients (off-takers) demand due to challenges related to estimating/optimizing yields from small-scale farming, where there is little or no digital footprint for a farm.
\item Mechanization service providers (i.e., tractor service providers) collectively indicated they do not have any clear visibility and insights into the activities performed by their fleets. 
That, for example, affecting their ability to understand 
when are tractors needed, how much being cultivated by each tractor, reputation or credibility of an operator, what was the quality of operator activity, when should tractors be serviced, are tractors in farming vicinity, how many more tractors to buy, etc.

\item Financial intuitions rarely extend any financial services (e.g., lending/credit) to farmers and fleet owners as means of return of investment especially due to lack of trusted and clear visibility on what is going on a farm and how a tractor being utilized. 
Note that fleet owners need to relay utilization (measure of profitability) of their fleet to their financial institution for affordable banking services. However, trust (as a result of lack of end-to-end visibility) continued to be one of the major issues among these entities.
For example, because of this trust issue today most of the tractor owners are subject to high interest rates (often 100\% collateral) in developing countries.

\end{itemize}

\subsection{Synthesis of the pain points and why Blockchain}

Analysis of the pain points clearly outlined the need for a technological solution that provides trusted information visibility with accurate representations of the whole agricultural business process (agricultural value chain). The value chain includes the following three aspects. First, a means of providing trusted farm activity provenance by defining, capturing and monitoring all activities that revolve around the farm while at the same time incorporating all the participants in each activity. Second, a relationship schema with access control logic(ACL) for all the defined activities which map out the workflow of all the farm activities expected to be performed. Third, drawing value additions (insights) from the collected activity data for all the agricultural stakeholders while enforcing data integrity, privacy, and transparency. Due to the highly fragmented agricultural ecosystem, the use of the blockchain platform perfectly aligns to address the pain points described above by establishing and promoting a decentralized trust among all stakeholders in the ecosystem. Furthermore, using smart contracts to encode and control the whole value chain activity transactions assures trust and integrity by leveraging on blockchain capabilities (e.g, trust consensus mechanisms, transaction finality, and non-repudiation). All representatives in the workshop were receptive to the idea of using blockchain for trusted information visibility and they consented that blockchain-powered farm record solution will be useful in improving various farm-related operations and agricultural decisions in general.

We believe blockchain offers a standards based information sharing platform. More specifically, permissioned Hyperledger Fabric (HLF)\cite{DBLP:journals/corr/abs-1801-10228}  blockchain platform provides decentralized  and  distributed  platform  that  is  useful  in  storing sensitive transactions  aimed  at  improving  trust,  transparency  and  integrity among the participants in a fragmented ecosystem (e.g. agriculture). In the agriculture domain there are already successful examples of large private corporations either trialling or mandating that their consumers will be able to scan codes on packaged products and be able to trace the travel of the foodstuff in their hand from its source in a farm, all through the supply chain, to its destination in a shop shelf (colloquially called farm-to-fork) enabled by blockchain. One example is IBM Food Trust \cite{ibmfoodtrust} which is a solution targeted at sharing certification and enabling traceability of food products being used by Walmart and other retailers.

Efforts to streamline financial transactions in agricultural ecosystems are also underway. Private sector examples in the trading space include LDC and ING \cite{coindesk} partnering to trial on soybean trade platform and the ADM, Bunge, Cargill, Louis Dreyfuss consortium considering wider efficiencies in global agricultural commodity trades \cite{Reuters}. Additional use cases include technology platforms that have been developed to establish digital marketplaces for agricultural inputs and direct-to-consumer access for growers.

\section{Use Case Scenario}
\label{sec:use-case-scenario}
\begin{figure*}[t]
	\centering
	\includegraphics[width=1\ScaleIfNeeded]{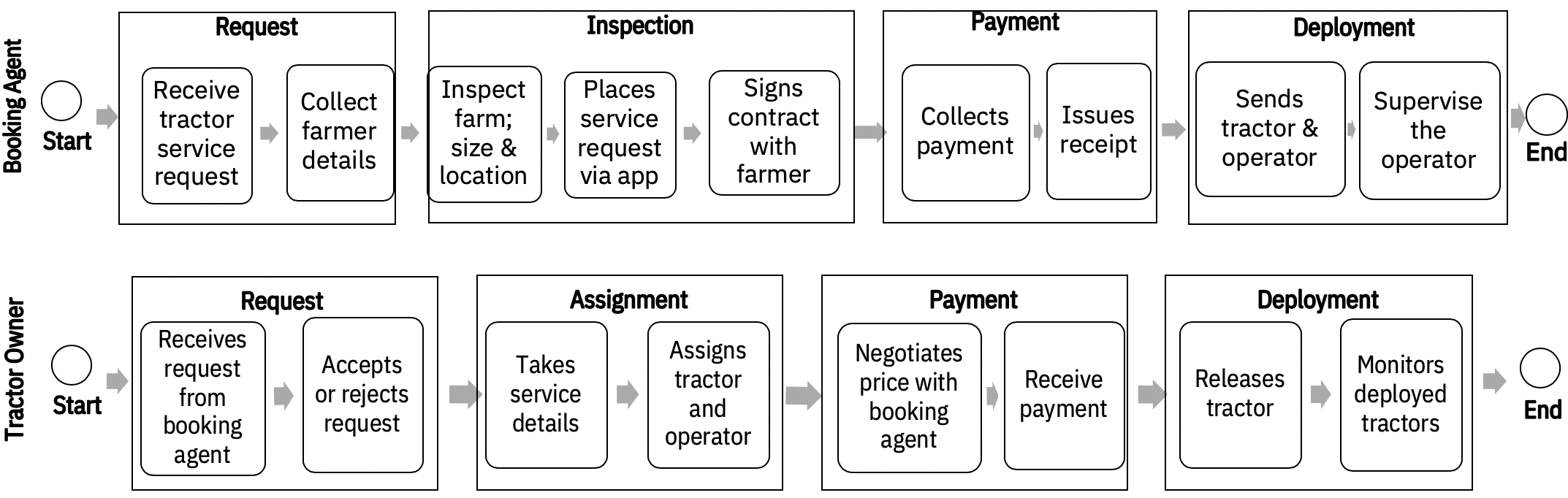}
	\caption{Simplified business process (workflow) for farm service request}   
    \label{fig:workflow}
\end{figure*}

For the remainder of this paper, we use the following use-case as a demonstration of the work. 

We are working with a partner (the ABC\footnote{We refer our partner the \say{ABC} as we are asked to protect its real name.}) who facilitates farm mechanization and data services for small-scale farming ecosystem using their platform. 
One of ABC's capabilities comes with a mobile booking application where farmers can request for mechanization services (e.g., ploughing). This can be done either the farm directly requests the service or work with a farm booking agent who submits the service request to ABC's platform on behalf of the farmer. This involves the agent walking around the farm to estimate the boundary of the farm while recording the geo-coordinates using ABC's booking application. Along with the boundary information, the agent also records the primary and secondary crop, tractor service type and some contextual data about the farmer and the farm. This request is termed as a booking request within the ABC platform and is then forwarded to the Scheduling app, which is used by tractor fleet managers to aid them in fleet and resource management. 

All requests are grouped into working clusters based on their location, farm size, type of skill needed. Each cluster is then paired with a suitable tractor-operator pair. The pairing process recommends, by taking into consideration various factors (e.g., skills, prior experience, etc.), a suitable tractor operator. With the pairing complete, all the requests in the cluster are then scheduled for the appropriate date and time where the soil in the farm is in the right condition for the requested service. Depending on the type of the service, additional process may apply. For example, if the service request is ploughing, farm soil condition may need to be assessed ---e.g., verifying recommended amount of moisture in the soil for the farm using weather data for optimum utilization of the tractor.  
%

The above estimated boundary information is key to determine the payment amount for the requested tractor service. Interestingly, the   recording of the geo-coordinates often gets over/underestimated to the advantage of the agent who often colludes with a farmer and a tractor operator without the knowledge of the fleet owner and/or a financier. This is just one of the many trust and transparency problems in this ecosystem. 

 Figure \ref{fig:workflow} is an excerpt of a simplified business process we mapped for the above use case. 
In the reminder of this paper, using this business process, we will formalize and digitize initial set of these relationships and interactions in a trustworthy and interoperable manner to ultimately enable the sharing of verified data and services within the small-scale farming ecosystem.




\section{ADW Design} 
\label{sec:systemoverview}

\begin{figure}[t]
 	\centering
 	\includegraphics[width=1\ScaleIfNeeded]{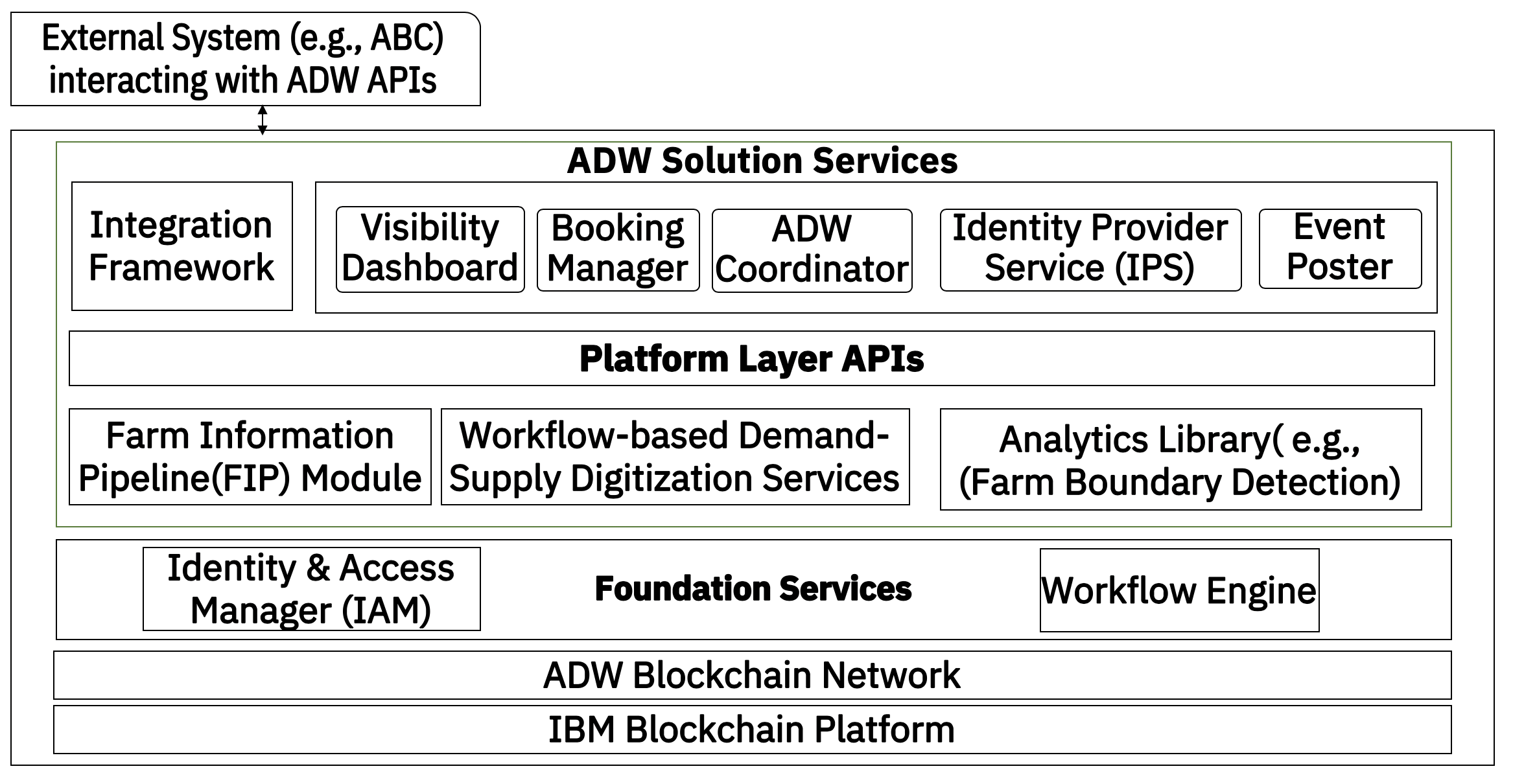}
     \caption{High-level view of the system components.} 
     \label{fig:system_overview_v4}
 \end{figure}

As shown in Fig. \ref{fig:system_overview_v4}, ADW platform layer comprises {Farm Information Pipeline} (Data), Workflow-based demand-supply digitization service (Farm Process),  and Analytics library (Data and Service). 
In the next section, we discuss each of these components in detail.

\subsection{Digitize farm demand-supply}

The ADW is designed to 
digitize and formalize the interactions of the small-scale ecosystem (consists of farmers, extension officers,  tractor owners and operators, dealers/technicians, and financial institutions) using smart contracts and blockchain network. The system enables a number of utilities pertaining to farm digitization and formalization.

We provide a tamper-proof definition of demand-side and supply-side workflows. These workflows are defined based on our extensive design thinking sessions and formal business process mapping of the agriculture value chains.The workflows map all \emph{farm} activities with different actors associated with complete farm activity provenance. Note that we represent each farm as an asset in the blockchain network using generic configurable workflows.
The activities defined in the workflow (e.g., create a booking for a farm service such as ploughing, add associated events, etc) provide a defined set of sequential actions and roles (of users) responsible for each of the actions that form part of a complete activity in the farm in context. Smart contracts (i.e., chaincodes as in HLF \cite{DBLP:journals/corr/abs-1801-10228}) are used to track and monitor each of the activity sequence against the farm, by enforcing workflow execution logic \cite{DBLP:conf/icbc2/BoreKMKORW19}. We also keep track of all workflow related approvals and digital confirmation of receipt of events at different steps (e.g., booking, confirmation of service delivery, invoice,  financing, etc.), documentation availability, tractor status, etc. 

HLF channels are used to enforce data privacy by isolating the storage of sensitive information (e.g., farm documents, activity data) for all participants for each agricultural co-operative business network. This ensures that the data is only distributed to specific participants in the channel by leveraging on blockchains chaincode logic, consensus and endorsement policies for data privacy and sharing restrictions. Each network has blockchain components as well as dedicated off-chain data storage mechanisms used to store de-identified data to satisfy privacy data requirements for organization. Data binding (linking) logic, farm workflow, and hash values are stored in the ledger to enforce access restrictions to de-identified EFR data stored off-chain. 

In each of the farm data collected, the workflow system explicitly outlines all participants responsible with their expected actions (e.g., create farm service booking) sequentially ordered. The ADW associates expected actions with read and write roles for all participants enrolled in agriculture business networks using the workflow logic implemented by chaincodes. Managing and controlling data associations in the ledger ensures that EFR data can be cross-referenced from different participants in a network. This includes granting authorized access to services and documents within the workflow associated with farm activities. In scenarios where there is a need to share existing data with other networks, ADW provides a mechanism where the workflow can be updated with the new data access request. Moreover, for every EFR data request, the workflow system serves as a gateway to check on the roles defined and provide access to on/off-chain data.  %
To support extensibility and heterogeneity across certain tiers (analytics, services) without entailing major re-work/dev effort, 
we provide integration endpoints with external system of engagements.

\subsection{Farm information pipeline}

 
 
\begin{figure*}[t]
	\centering
	\includegraphics[width=1\ScaleIfNeeded]{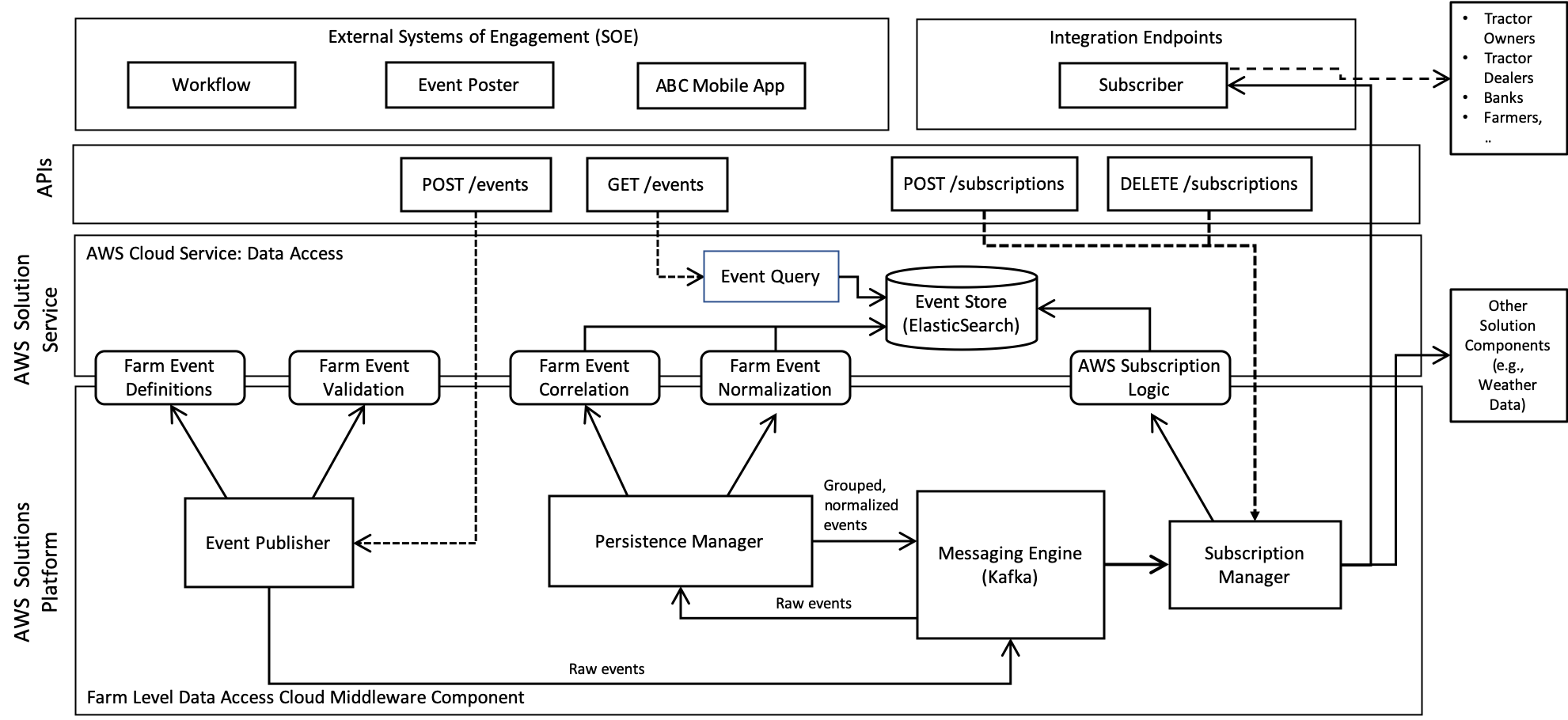}
	\caption{Farm Information Pipeline (FIP)}   
    \label{fig:fip_diagram}
\end{figure*}

The primary objective of farm information pipeline (FIP) is to ingest farm activities (events) from each System of Engagement (e.g., ABC's IoT system) and then provide secure storage of those farm activities.  All activities/events are persisted, where the persistence interval are set by policy. We used these farm activity data to accurately identify and estimate farm boundaries and acreage and then distribute those events to trusted organizations in a secure and timely manner.


Figure \ref{fig:fip_diagram} is the ADW's FIP process to stream data sources corresponding to farm activities. 
The pipeline process starts by mapping the geographical attributes for each farm, most importantly, the exact center location of all farms of interest. To do that, we processed day-stamped GPS logs from a large network of active low-horsepower (low-cost) in-field tractors, each log represents a day of plowing activity linked to a particular tractor.  The farm event publisher service defines farm events that cater to different events expected from IoT sensors (e.g., sensors attached to tractors). The definitions are then used to validate the events as they are received from the IoT sensors after which they are correlated and normalized with other events based on a farm record in EFR and its associated activity (e.g., booking requests). In addition, FIP provides a subscription logic for event queries following farm workflows and enforced by the blockchain. Remote sense data (e.g., remote sensing images stored in \cite{IBMPAIRSSERVICES}) and weather data corresponding to each farm are also pulled and integrated with the farm activity data. 

We enabled any third party system to publish farm or tractor-level events throughout the farm mechanization demand-supply process. Farm/tractor level information is accessible to ecosystem participants in real-time (farm boundary, estimated acreage, serviced area, etc.).  Through FIP and machine learning algorithms, difficult to obtain pieces of farm and tractor level information is provided to allow for intelligent decision making (e.g., tractor job completion to trigger payment). FIP also provides a publication and subscription service for farm and tractor related events. It can also consume current electronic data interchange (EDI) type messages, convert them to FIP format and publish them. 


\subsection{Data analytic services}

Using blockchain verified EFR and machine learning algorithms, ADW makes available farm and/or tractor level visibility information and insights that are accessible to ecosystem participants in real-time.  Such difficult to obtain pieces of farm-level insights (for small-scale farming ecosystem) is provided to allow for intelligent decision making. 
Examples of machine learning services we have enabled currently include region based tractor allocation, farm service schedule recommender engine (e.g., determining the appropriate date to plough a farm), automated field boundary identification and acreage estimation (see an example in Figure \ref{fig:farm_activity}), and credit scoring for fleet owners. 
Tractor-level insights are also generated using machine learning, which can be visualized and tracked (see an example in Figure \ref{fig:dashboard}). Examples of tractor-level insights include the ability to visualize the utilization pattern of a tractor, the total farms served to-date by this tractor, revenue generated out of a tractor, as well as insight about when a particular tractor should not be used along with any other anomalous behavior associated with the tractor.

The lifecycle of these models are managed by the blockchain system and any interaction with these models are recorded on the blockchain. Every action on a model is recorded as a blockchain event or sequence of events for transparency and immutability. Before a model is executed, the model file is verified by executing smart contracts.
\section{Implementation }

\begin{figure*}
    \centering
    \begin{subfigure}[b]{0.48\textwidth}
       \includegraphics[width=1\ScaleIfNeeded]{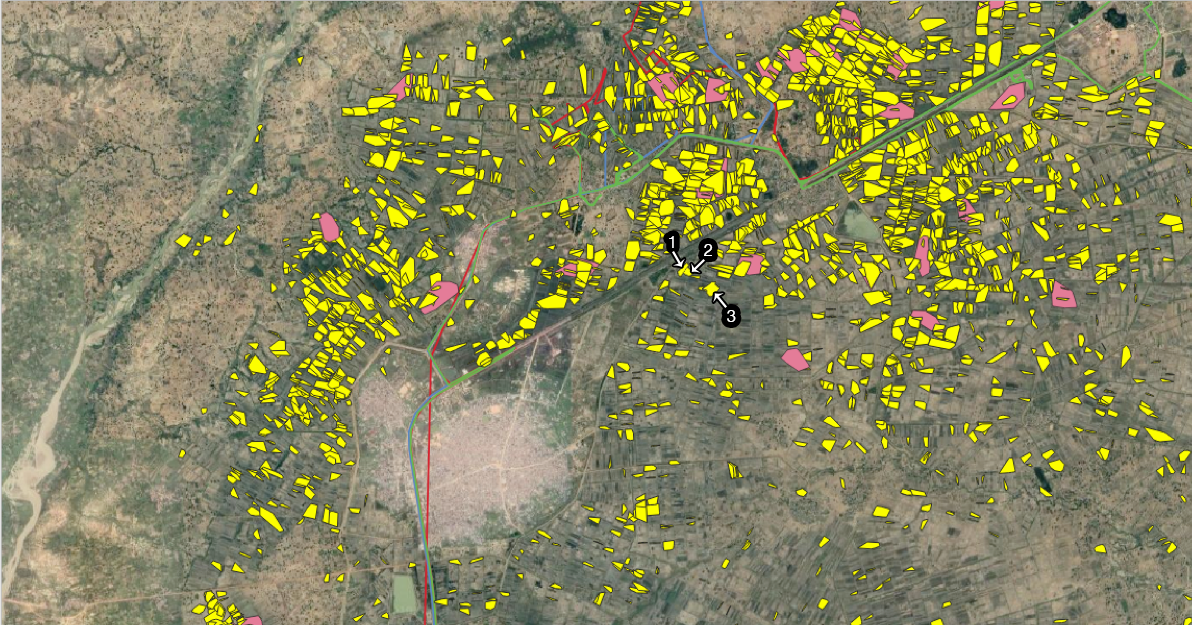}
   	\caption{Blockchain validated farms their boundaries are computed using machine learning models. Farms 1, 2 and 3 form the cluster shown in Figure \ref{fig:dashboard}. }
		\label{fig:farm_activity}
    \end{subfigure}
    ~ 
    \begin{subfigure}[b]{0.48\textwidth}
       \includegraphics[width=1\ScaleIfNeeded]{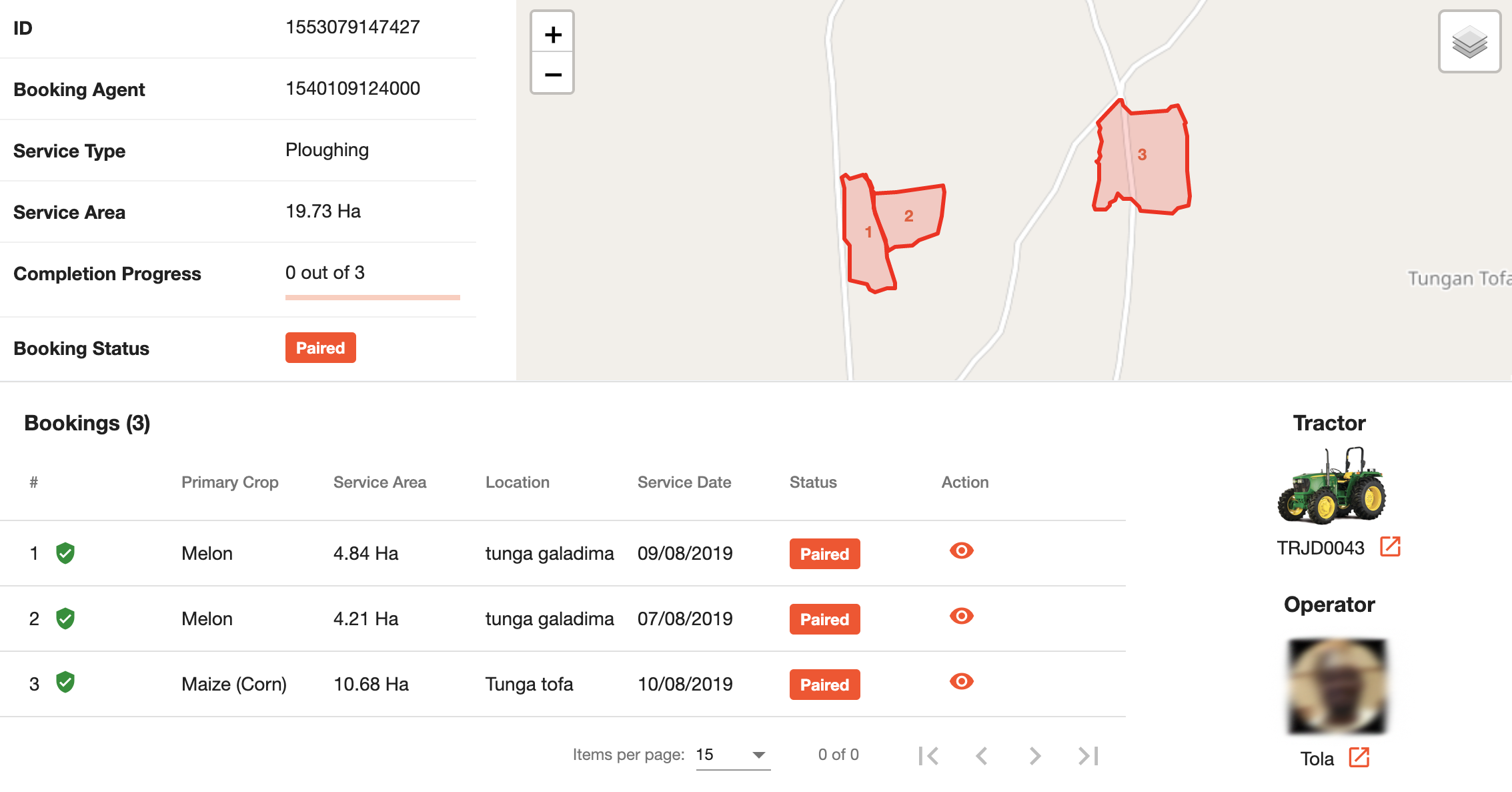}
    \caption{Visibility dashboard showing a cluster of three blockchain validated farms for a single mechanization service as seen by a fleet manager.}
        \label{fig:dashboard}
    \end{subfigure}
    ~ 
    \caption{(a) Real-world small-scale farms in Northern Nigeria and (b) a Fleet Manager visibility dashboard on the ADW.}\label{fig:visual_aws}
\end{figure*}

ADW is implemented as a suite of cloud foundry micro-services and chaincodes are implemented on HLF blockchain.
The actors defined in the farm workflow are managed by an Identify \& Access Manager (IAM), which builds upon the HLF's membership service providers. IAM's main role is to provide access control capability to farm activity data using blockchains secure authentication mechanisms. It also encodes and enforces data (including documents) sharing rules in the workflow. Each request (e.g., data views or post) received by ADW is uniquely authenticated using tokens generated by HLF's Identity Provider Service (IPS) ---an OpenID connect compliant service \cite{DBLP:conf/secsr/LiMC19}. IPS uses IAM to register and enroll users in the network for enrollment certificates to be used in authenticating users in the network and also sign valid transactions to the ledger. For our implementation, we also leveraged IPS to manage Single Sign-On (SSO) service providers from 3rd-party components integrated with the ADW. 

 
The event bus for FIP, manages the event logging and event-based interaction, is implemented using Kafka \cite{kafka}. Our implementation follows the \say{Event Sourcing} design pattern \cite{Pacheco:2018:MPB:3208677}, where Kafka is the frontend for all the event logging and event interaction management through pub-sub mechanism. 
For complete farm demand-supply digitization provenance, we implemented demand-side and supply-side workflows with respective participants for each activity using JSON encoded farm-based workflow. To manage and execute workflows, we extended the workflow engine as discussed in \cite{DBLP:conf/icbc2/BoreKMKORW19}. This workflow engine has inbuilt multi-chaincodes for workflow execution while enforcing data access roles for all participants of the network. And, using hash values stored in the ledger, we compare and validate data related to farm activities. 
Data (and model) accesses are verified and fine-grained data (and model) level ACLs are enforced for the off-chain IoT data (Postgres). Data are persisted, versioned, encrypted and indexed for operations, analytics and reporting.



We implemented an in-built data syncer to manage the transfer of daily tractor activities corresponding to booking events from ABC to ADW blockchain network via FIP endpoints following the use-case described in Section \ref{sec:use-case-scenario}. Data privacy was enforced by syncing users correlation id defined by access control logic and stripping off any personal identification information in the data. Overall, approximately over 100,000 farm-level activity events were resolved from data collected on a platoon of instrumented tractors and processed by the FIP. The farm-level activity events include groups of one or more small-holder farms with some being revisited multiple times. These events represented 19,799 small-scale farms for which we were able to accurately detect farm boundaries for 12,718 farms using ADW automatic farm boundary detection module. The average size of a farm is estimated to be 2 hectares 
derived from detected farm boundaries.  On average, 7 farms are serviced per day with the bulk of the activities taking place during the ploughing seasons in preparation for planting.


Figure \ref{fig:farm_activity} shows examples of some real digitized blockchain-enabled farms (as maintained in the EFR), each farm boundary has been identified and characterized using machine learning models. As shown in the figure, the transactions and data from both the farm supply-demand system and FIP are represented visually to reflect actual tractor trace route information with real-time satellite images (avaiable in \cite{IBMPAIRSSERVICES}) to create a visual representation of EFR blockchain records as shown in Figure \ref{fig:visual_aws}. We also implemented algorithms to cluster and assign farm service jobs to networks of tractor service providers.   External APIs centered around farm  management and exposing computed analytics on an integrated dashboard as shown in Figure \ref{fig:dashboard}. This dashboard provides  access to the data and key mechanization and agriculture analytics necessary for making informed agribusiness decisions.

\section{Performance Evaluation of ADW}
\label{sec:experiment}


Before deploying the ADW, we have performed extensive performance evaluations of the overall system. In this section, we particularly focus on the evaluation of ADW's blockchain components. The Linux foundation through its Hyperledger open source collaborative effort proposes several key performance considerations for designing valid reproducible blockchain systems in \cite{HYPERLEDGER}. One of their suggestions is the evaluation of blockchain latency and throughput performance aspects. This includes evaluating various network setups by considering multiple transactions characteristics and varying workload generation. We evaluated these two common performance metrics for our system: (i) by altering the total number of transaction subjected to ADW per second in order to determine the maximum number of transactions that the system can commit to the ledger successfully in a second; and then, (ii) by observing the impact of different number of transactions grouped as a block size in the ledger to assess the overall ADW performance.




Thus, for \textit{throughput} we evaluate the total number of transactions per second that are successfully processed by the ADW blockchain network. And, for \textit{latency}, we evaluate the total time taken when a request is sent and confirmed in the network to the time when our blockchain client receives a response back. We  used  the  Hyperledger Caliper \cite{caliper} benchmarking tool supported by the Linux foundation for evaluating hyperledger fabric performance indicators such as throughput and latency. 

All experiment requests were subjected to ADW workflow engine which is responsible for committing the transactions to the ledger. Evaluation of the impact of integration of ADW blockchain component with other solution services (e.g., booking manager) did not degrade overall performance of the system. This is because the different services (e.g., solution services, farm information pipeline, analytics services, etc.) are designed as independent micro-services that perform resource intensive operations while leveraging and/or interacting with the ADW blockchain network.



\begin{table}[t]
\caption{ Experiment configuration}
\label{tab:configuration}
\begin{tabular}{ |p{2.5cm}| |p{5.5cm} |  }
   \hline
 \textbf{Parameters}  & \textbf{Values} \\
   \hline

 Hyperledger  & Fabric v1.4.1 \\

 StateDB database	& GoLevelDB \\
 Transaction send rate & 20, 40, 60, 80, 100, 120, 140, 160, 180,200\\
 Block size	& 10, 30,  50,  70\\
 Block timeout	& 500ms	\\
 Number of channels &	1	\\
  \hline
 \end{tabular}
 \end{table}

Table~\ref{tab:configuration} shows our initial test configurations for ADW experiment setup. Experiment test load was generated by 25 simulated blockchain clients each using a constant send rate increment of 20 transactions(tx)/second(sec) starting from 20 tx/sec to 200 tx/sec with a target of 1000 tx/sec for each increment and increase of block size as shown in the table. The blockchain network included three organizations   
each with one peer and one communication channel. 

Figure \ref{fig:throughput} shows throughput measurements with a maximum load of 200 tx/sec. The results shows invoke workloads where throughput increases linearly as expected and starts to saturate at around 110 tx/sec. Average latency also increased steadily as send rate increased to approximately 5 seconds as it approached saturation point. For input transaction rate closer to ADW saturation point for example when input rate was 200 tx/sec the latency was higher due to increase of block creation time and transactions had delays at the orderer node. The is due to increase in the number of transactions waiting in HLF queue during block creation and validation which affected commit latency as described by \cite{cvcbt}. A slightly higher latency of approximately 1 sec is observed especially with a higher block size (e.g., 70) at 80 tx/sec. The reason for the higher latency is due to a higher block creation time caused by a higher block size leading to transactions waiting a little longer at the orderer node.


Latency decreased with an increase of block size from 10 to 70 especially when input rate was higher than ADW saturation point(e.g. 120 to 200 tx/sec) due to less time taken to validate and commit each block size for all blocks available. ADW average throughput results of 110 tx/sec is strikingly lower from other HLF related implementations which report a higher throughput ranging from 1000 tx/sec to 3500 tx/sec \cite{IBM,DBLP:conf/cvcbt/BaligaSVPKC18,cvcbt}. This major difference is attributed to the complex business logic implemented by multiple chaincodes used to digitize farm demand-supply activities encoded as farm workflows (e.g., automatic workflow execution to next action logic and mapping out participants roles for all activities in the workflow). To address low performance results,  performance optimization metrics and guidelines which include introduction of membership service provider cache and bulk read/write during multiversion concurrency control validation which are adequately explained by \cite{DBLP:conf/cvcbt/BaligaSVPKC18} and \cite{cvcbt} will be implemented in the future version. In addition, our future version will leverage on performance improvements in HLF v2.0 \cite{hyperledgerv2.0} which include parallelized transaction during commit phase, improved transactions cache and asynchronous write block processing to effectively manage multiple organizations and peers in large-scale deployments.




\begin{figure}[t]
	\centering
		\includegraphics[width=1\ScaleIfNeeded]{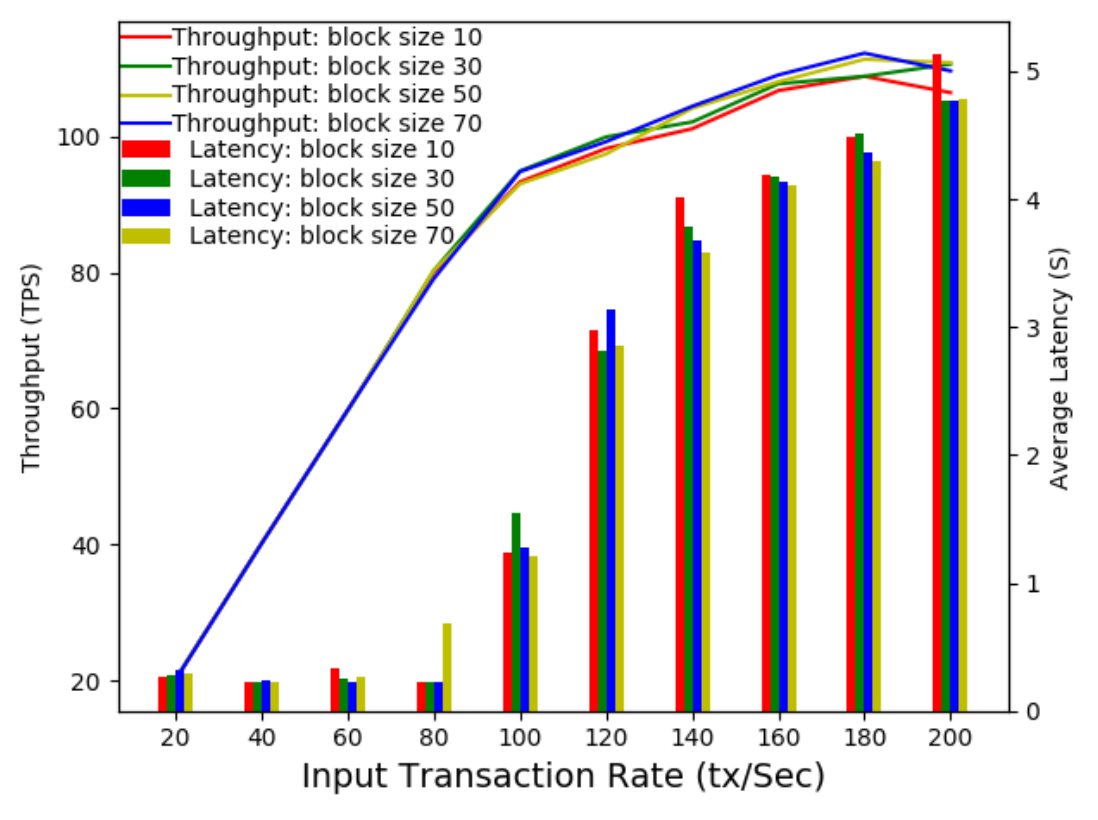}
    \caption{Throughput and average latency performance of ADW.}
    \label{fig:throughput}
\end{figure}




\section{Real-world deployment}

Below we discuss the pilot deployment of ADW and report initial qualitative feedback we received from the ecosystem. 
\subsection{Deployment}
We deployed ADW as a stack of cloud foundry(CF) apps (micro-services)  
connected to ADW's HLF blockchain network.   Our ADW inbuilt multi-org, multi-users and multi-channel configuration enabled the ADW system to handle multiple requests from various agricultural networks concurrently. ADW's blockchain network included two  blockchain  peers/nodes per each of the 3 organization clusters on IBM's enterprise blockchain platform.
Note that our deployment mimics the typical structure of an agriculture network using the multi-tenant network. 

Participating small-scale farming ecosystem clusters each host and manage a stack of cloud foundry(CF) apps. Each stack includes the ADW blockchain platform components as well as dedicated blockchain-managed FIP for that stack. Any sensitive information including invoices are distributed only to those stacks participating in a channel. 
%
Therefore, the first cluster (the booking agent/farmer cluster) was configured to manage, collect, and authenticate all booking requests for farm services from booking agents. We configured the second cluster for managing tractor owners with operators and enable interaction with trusted data value insights enabled on blockchain. This is useful for information retrieval as part of information visibility (e.g. reporting on tractor and operator activity, performance, the maintenance needs of a tractor, and requests for tractor services). The third  cluster is configured to manage on-boarding and access  of agricultural  networks  ---e.g., ABC to transmit  IoT data to ADW's FIP which enables authorized access to myriad agricultural services and agribusiness data insights from machine learning models. 


The CF apps/services for each organization include booking manager that has inbuilt coordinator service that receives data from external requests from ABC, centralized analytic services that expose data insights and identity provider service that manages default authentication mechanisms as well as using 3rd party authentication (SSO) providers from the external ABC service connecting to the ADW. These services were configured to use {Postgres} as an off-chain document store.
Our deployment supports \say{bring-your-own doc-store} (\say{via pluggable storage device}) with highly customized document store architecture.

\subsection{Preliminary Feedback}

Currently, ADW is in pilot use since for the last 3 months in Nigeria (users and organizations from Malawi, Kenya, Thailand and Senegal also signed up).  The pilot test was designed to evaluate the following key factors: a) use of the digitized farm activities using ADW to establish trusted electronic farm records visibility b) the impact and experience of the ADW to tractor owners in their operations and other stakeholders including credit from financial institutions. 

Prior to the pilot kick-off, we have conducted extensive training of tractor owners and farm agents in Kenya, Nigeria, Malawi, Thailand, and Senegal. The choice of these regions was due to the presence of ABC farm mechanization services which enabled us to access IoT data from tractors. The training of tractor owners included how they will access and use the ADW dashboard to manage bookings trends and summary, insights on their tractor revenue projections, insights on their fleet capacity operations which include service area, tractor utilization and fuel consumption insights. We also conducted interviews with representatives from financial institutions that provide financing to tractor owners to understand what they consider before offering lending or loans. They also evaluated the financier dashboards and provided feedback on how these would help them in making financing decisions.

We received constructive feedback and suggestions from the different stakeholders who were part of the training and pilot program. For example, one tractor fleet owner stated that, 
\begin{quoting}
 \say{... great to have such a system. I can see how ADW enables me to monitor the utilization of my fleets and activities in a more trusted way. I can also see how projected revenue I would enjoy over time based on the evidence from the data.}
 \end{quoting}
 A financial institution representative stated that,
 \begin{quoting}
 \say{... with ADW platform we can now consider to offer 80\% financing on tractor loans and ... }
 
 \end{quoting}
A group of booking agents who also representing small-scale farmers excitedly reported that,
\begin{quoting}
 \say{... ADW provided opportunities for farmer centered services which include agronomy services targeting small-scale farmers at different regions. It is clear that financial institutions can use insights from EFR to finance small-scale farming.}
 \end{quoting}

\section{Conclusion and future work}
\label{sec:conclusion}


In this paper, we presented ADW --- a blockchain based electronic field records (EFR) system to formalize and digitally represent farms based on data collected using various technologies such as mobile-based data collection app (including longitudinal data extraction), weather remote sensing, and IoT sensors (tractors), etc. 
ADW was designed in tight collaboration with the small-scale farming ecosystem whom we conducted extensive workshops and field visits (including interviewing various representatives from the ecosystem).
We ingested over one hundred thousand small-scale farms activity events and developed machine learning models to create trusted electronic field records (EFR) and automated farm boundary detection. All these are configured with the ADW blockchain network.

While the current pilot is progressing as expected, we already started noticing complex issues as the network grows with farm cooperative networks.  This requires us to design better mechanisms for orgs onboarding and for data and transaction privacy (via multichannel rules). Pre-determined rules (e.g., partitioning logic as common knowledge) can be defined for the farm network participant list to determine which channel it belongs to.  Our current performance evaluation is low (for a known reason ---i.e., poorly resource-constrained connections) compared with benchmarks. Thus, further optimization of ADW and performance evaluations (including bulk read/write during multi-version concurrency control for validation and commit phases optimizations) are needed.  We will continue enhancing EFR models with additional data (remote sensing and weather) and services such as yield forecast, demand-and-supply management, etc.  Based on the findings  from the initial pilot tests, we will optimize and deploy ADW in multiple countries (as requested by ABC and several other players in the small-scale farming ecosystem). Finally, we plan to assess the socio-economic impacts of ADW by working with usability researchers and domain experts. 

\bibliography{bibliography}
\bibliographystyle{IEEEtran}

\end{document}